\documentclass[prl,nofootinbib,floats,aps,twocolumn,tightenlines,superscriptaddress,floatfix]{revtex4-1}

\usepackage{epsf,epsfig,amsmath,amssymb,dsfont}
\usepackage{amsthm,mathrsfs} %mathematical symbols
\usepackage{graphicx} %includes images
\usepackage{multirow}
\usepackage{float}
\usepackage{colortbl}
\usepackage[table]{xcolor}
\usepackage{varioref}
\usepackage{slashed}
\usepackage{float}
\usepackage{hyperref}
\usepackage{longtable}
\usepackage{array}
\usepackage{mathtools}

\newcommand{\be}{\begin{equation}}
\newcommand{\ee}{\end{equation}}
\newcommand{\baln}{\begin{align}}
\newcommand{\ealn}{\end{align}}
\newcommand{\ben}{\begin{equation*}}
\newcommand{\een}{\end{equation*}}

\long\def\symbolfootnote[#1]#2{\begingroup%
\def\thefootnote{\fnsymbol{footnote}}\footnote[#1]{#2}\endgroup}

\newcommand{\fr}{\frac}

\begin{document}

\title{%Inertial 
Low energy signatures of  nonlocal field theories}

\author{Alessio Belenchia}
\affiliation{SISSA - International School for Advanced Studies, Via Bonomea 265, 34136 Trieste, Italy.}
\affiliation{INFN, Sezione di Trieste, Trieste, Italy.}\author{Dionigi M. T. Benincasa}
\affiliation{SISSA - International School for Advanced Studies, Via Bonomea 265, 34136 Trieste, Italy.}
\affiliation{INFN, Sezione di Trieste, Trieste, Italy.}
\author{Eduardo Mart\'{i}n-Mart\'{i}nez}
\affiliation{Institute for Quantum Computing, University of Waterloo, Waterloo, Ontario, N2L 3G1, Canada}
\affiliation{Department of Applied Mathematics, University of Waterloo, Waterloo, Ontario, N2L 3G1, Canada}
\affiliation{Perimeter Institute for Theoretical Physics, 31 Caroline St N, Waterloo, Ontario, N2L 2Y5, Canada}
\author{Mehdi Saravani}
\affiliation{Perimeter Institute for Theoretical Physics, 31 Caroline St N, Waterloo, Ontario, N2L 2Y5, Canada}
\affiliation{School of Mathematical Sciences, University of Nottingham, University Park, Nottingham, NG7 2RD, UK}

%\emailAdd{abelen@sissa.it}
%\emailAdd{dionigi.benincasa@sissa.it}

\begin{abstract}

The response of inertial particle detectors coupled to a scalar field satisfying nonlocal dynamics
described by nonanalytic functions of the d'Alembertian operator $\Box$ is studied. We show that spontaneous emission processes of a low energy particle detector are very sensitive to high-energy nonlocality scales. This allows us to suggest a nuclear physics experiment ($\sim$ MeV  energy scales) that outperforms the sensitivity of LHC experiments by many orders of magnitude. This may have implications for the falsifiability of theoretical proposals of quantum gravity.
\end{abstract}

\maketitle
\flushbottom

\section{INTRODUCTION}%{\bf \textit{Introduction.-}} 
Quantum field theories with nonlocal dynamics were originally studied in the 1950s and 1960s 
with the goal of sidestepping the infinities of local interacting QFTs~\cite{Pais:1950za}.%Yukawa1950 
With the advent of Wilson's understanding of renormalization and the birth of the Standard Model however,
these attempts were by and large abandoned only to be revived in the last two decades, 
%with interest in nonlocal field theories back on the rise, 
mainly because they seem to emerge ubiquitously in models of quantum gravity~\cite{Sorkin:2007qi,Moeller:2002vx,Gambini:2014kba}, 
and also because they provide examples of consistent, 
renormalizable theories of gravity~\cite{Modesto:2011aa}. 

Nonlocal field theories are simply defined as field theories whose equations of motion
have an infinite number of derivatives.
For example, the equations of motion for a nonlocal free massless scalar field 
can be written in the form $
f(\Box)\phi(x) = 0$, 
where $f$ is some (nonpolynomial) function of $\Box$. %({\bf DB: maybe a few words about
%field redefinitions for noninteracting NLQFTs?)}.
These theories can be subdivided into two subclasses: a)  those defined by entire analytic  functions $f$ 
and, b) those with nonanalytic $f$s.  
In both cases the nonlocality of the theory can lead to a much improved ultraviolet (UV) behaviour
of the propagators~\cite{Efimov1967,Aslanbeigi:2014zva}, which is the reason why these theories were originally studied. The qualitative behaviour of the two subclasses a) and b) is, however, radically different. The underlying reason for this is that unlike an entire analytic function, nonanalytic functions
contain a branch cut, i.e. a 1-dimensional subspace of the complex plane where the function has a 
discontinuity. In the Green function this branch cut corresponds to a continuum of massive modes, even though
the original field itself is massless. Note that this is very similar to what happens to the Green function of local interacting QFTs~\cite{Weinberg:2005}. 
As we will discuss below, the presence of this continuum of massive modes modifies all $n$-point
functions of the theory, thus giving rise to nontrivial modifications to many physical observables.

Much of the early literature on nonlocal field theories was devoted to understanding 
properties such as stability and unitarity~\cite{Gomis:2000aa,AlvarezGaume:2001ka,Barci:1997xy}.
%, in particular for the case of analytic $f$s
%Oxman,Noncommutiative,SFT.
Recently however, there
has been considerable interest in the extraction of phenomenological consequences of this kind of nonlocality, for both analytic and nonanalytic $f$s~\cite{Biswas:2014yia,Belenchia:2015ake,Saravani:2015rva,Barnaby:2008fk}. 

Ideally, one would like to find experimentally accessible signatures of nonlocality so that its existence can be put to the test. However, if such a scale is assumed to be near the Planck scale, finding an experimental setup in which the nonlocal features of the theory can be seen becomes extremely challenging.

The fact that the low energy behaviour of particle detectors is sensitive to high-energy effects was recently pointed out by Kajuri~\cite{Polymer}, and Louko and Husain~\cite{ViqarJorma}. They showed that some features of low energy particle detectors can be sensitive to violations of Lorenz invariance at high energies. For example, in~\cite{ViqarJorma} it is shown that polymer quantization (motivated by loop quantum gravity) may induce a Lorentz violation at high energies that is perceived by low energy detectors (below current ion collider energy scales). 
More concretely, they found low energy Lorentz violations in the response of atoms modelled as Unruh-DeWitt detectors (which capture the features of the atom-light interactions~\cite{Martin-Martinez2013,Alvaro}) for a general family of quantum fields with modified dispersion relations at high (Planckian) energies. 

In contrast to~\cite{ViqarJorma}, we will consider nonlocal theories with nonanalytic $f$s that, crucially, preserve Lorentz Invariance (LI). It should be noted that LI violations are strictly constrained by various experimental observations, making theories that preserve LI
%, while simultaneously introducing new effects,
particularly appealing~\cite{Liberati:2013xla,Mattingly:2005re}. 

In this paper we show that the existence of a nonlocality scale in a scalar field theory has phenomenological consequences on the low energy behaviour of particle detectors. In particular, we study how the existence of a nonlocality scale 
%in the field theory 
influences the spontaneous emission of an atomic species: a very well understood and easy to test experimental setup. We will show that it is possible, in principle, to devise a finite-time low energy  experiment  with a resolution similar to that of particle collider setups.

%Unless otherwise stated, we use natural units \mbox{$c=\hbar=1$.}

\section{NONLOCAL DYNAMICS}%{\bf \textit{Nonlocal Dynamics.-}}
We study a real scalar field obeying a special class of nonlocal dynamics 
given by real, retarded, Poincar\'e invariant wave operators, $\widetilde{\Box}\coloneqq f(\Box)$. The retarded nature of these operators implies that $f$ is nonanalytic \cite{dominguez1979laplace}. Interest
in this particular kind of operators can be traced back to the original construction of R. Sorkin 
of a d'Alembertian operator on a 2 dimensional causal set~\cite{Sorkin:2007qi}. His results were
then extended to higher dimensions in~\cite{Benincasa:2010aa,Dowker:2013vl}, finally culminating in a comprehensive
study of all possible generalizations in all dimensions in~\cite{Aslanbeigi:2014zva}.
The operators $\widetilde{\Box}$ depend on a nonlocality scale $l_n$,
thus consistency with the local d'Alembertian requires that $\widetilde{\Box}\rightarrow\Box$ in
the limit $l_n\rightarrow0$ (for further details see~\cite{Aslanbeigi:2014zva}). 
A spectral analysis of these operators~\cite{Aslanbeigi:2014zva} reveals that as a function
of spacetime momenta the operators depend on both $k^2$ {\em and}  sgn($k^0$), i.e.
\be
\widetilde{\Box}e^{ik\cdot x}=B(\text{sgn}(k^0),k^2)e^{ik\cdot x}.
\ee
The function $B$ possesses a branch cut along $k^2\le0$ that represents a continuum of massive modes, much like those present in interacting local quantum field theories, except for the lack of a mass gap.
%The generic root/singular structure of $B$ in the complex $k^2$ plane  
%is shown in Figure \ref{poles}. 
%\begin{figure}[tbp]
%\begin{center}
%\includegraphics[scale=0.38]{polestructure.pdf}
%\caption{\small{}} 
%\label{poles}
%\end{center}
%\end{figure}
%The zeros of $B$ away from $k^2=0$ signify 
%potential classical instabilities, and always appear in complex conjugate pairs
%\cite{Belenchia2014}, 

Free and interacting scalar quantum field theories based on this family of dynamics 
have been constructed using different quantization schemes \cite{Belenchia:2014fda,Saravani:2015rva}, all of which lead to 
the same quantum theory, at least at the free level. In particular, the Wightman
function $D^{(+)}(x,y):=\langle0|\phi(x)\phi(y)|0\rangle$ for the free theory is given by \cite{Saravani:2015rva}
\begin{align}
D^{(+)}(x-y) = &\int \fr{d^4k}{(2\pi)^4} \widetilde{W}(k^2) e^{ik\cdot(x-y)},
\label{wightman}
\end{align}
where
\be
\widetilde{W}(k^2) = \fr{2\text{Im}(B)\theta(k^0)}{|B|^2}.
\label{discont} 
\ee
%In \eqref{wightman} we ignored the possible existence of complex mass modes. We will further comment on this in the outlook section. 
%a property that follows from the fact that 
%\be
%B(\text{sgn}(k^0),k^2)^*=B(-\text{sgn}(k^0),k^2).
%\ee
The Wightman function can be re-written as
\begin{align}\label{wight}
&D^{(+)}(x-y) = \int \fr{d^4k}{(2\pi)^4} 2\pi\theta(k^0)\delta(k^2)e^{ik\cdot(x-y)} \\
&+\int_0^\infty d\mu^2 \rho(\mu^2)\int \fr{d^4k}{(2\pi)^4} 2\pi\theta(k^0)\delta(k^2+\mu^2)e^{ik\cdot(x-y)}, \nonumber
\end{align}
where $2\pi \widetilde{\rho}(-k^2) = \widetilde{W}(k^2)$ and $\widetilde{\rho}(\mu^2)=\delta(\mu^2)+\rho(\mu^2)$.
One can see that $D^{(+)}$ is a sum of two parts, one is the standard Wightman function 
for a local massless scalar field, $D^{(+)}_0$, and the other is an integral over the Wightman function of a local 
massive field, $G^{(+)}_\mu$, weighted by the finite part of the discontinuity function, $\rho(\mu^2)$.

For every choice of $\widetilde{\Box}$ there corresponds a specific $\rho$. In this paper we are interested in two different kinds of d'Alembertians whose discontinuity functions are given by
\begin{align}\label{discfunc2d}
\rho(\mu^2)
%&:=i(f^{-1}(k^2+i0)-f^{-1}(k^2-i0))\nn
&=\lim_{\epsilon\rightarrow0^{+}}\fr{-2\,e^{l_n^2\mu^2/2}}{\mu^2} \fr{\Im[E_2(l_n^2(-\mu^2+i\epsilon)/2))]}{|E_2(-l_n^2\mu^2/2)|^2}.
\end{align}
and 
\be\label{2disc}
\rho(\mu^2)
%&:=i(f^{-1}(k^2+i0)-f^{-1}(k^2-i0))\nn
=l_n^2e^{-\alpha l_n^2\mu^2}.
\ee
where $\alpha$ is an order one numerical coefficient~\cite{Saravani:2015rva}. The former choice of $\rho$
can be shown to give rise to a stable interacting QFT~\cite{Saravani:2015rva}, while the latter is a much simpler function which captures all the fundamental features of \eqref{discfunc2d} (see~\cite{Aslanbeigi:2014zva,Saravani:2016enc}) and allows us to check that our results are largely independent of the specific form of the discontinuity function. 
Note that the asymptotic limit of the discontinuity function for small masses is given by $\rho(\mu^2)=l_n^2$~\cite{Saravani:2016enc}, 
while for large masses it is exponentially suppressed (see Appendix B of~\cite{Aslanbeigi:2014zva}).

\section{COUPLING THE FIELD TO A PARTICLE DETECTOR}%{\bf \textit{Coupling the field to a particle detector-}} 
The interaction of our nonlocal field with a two-level Unruh-DeWitt detector is described
by the interaction Hamiltonian $H=g\,  \chi(\tau/T)m(\tau)\phi[x(\tau)]$, where $g$ is a small coupling constant,
$m$ is the detector's monopole moment, and $x^\mu(\tau)$ are the detector's worldline 
coordinates parametrised by proper time $\tau$. We have included a switching function $\chi$ that controls the time dependence of the detector's coupling strength, and is strongly supported for a timescale $T$. This detector model captures the fundamental features of the light-matter interaction 
in the absence of angular momentum exchange~\cite{Martin-Martinez2013,Alvaro}.

The response function of an Unruh-Dewitt detector
is~\cite{Birrell1984,Sriramkumar:1994pb}  
\begin{align}
\mathcal{F}(\Omega, T)&= \int_{-\infty}^{\infty}\!\!\!\!
d\tau\int_{-\infty}^{\infty} \!\!\!\!d\tau' e^{-i\Omega\Delta\tau} D^{(+)}(\Delta\tau)\chi\Big(\frac{\tau}{T}\Big)\chi\Big(\frac{\tau'}{T}\!\Big),
% \chi(\tau,T)\chi(\tau',T),
%\int_{0}^{\infty}\rho(\mu^{2})\int_{-\infty}^{\infty}
%d\tau\int_{-\infty}^{\infty} d\tau' e^{-i\Omega\Delta\tau} G_{\mu}^{(+)}(\Delta\tau) \chi(\tau,T)\chi(\tau',T)\\ \nn
%&= \int_{0}^{\infty}d\mu^2\rho(\mu^{2}) \mathcal{F}_{\mu}(\Omega, T)
\label{transition_rate}
\end{align}
%Since the nonlocal Wightman function \eqref{wight} splits into the sum of
%a massless Wightman function and a contribution from a continuum of massive modes, 
where $\Omega$ is the energy difference between the two detector states and $\Delta\tau=\tau-\tau'$.
Using \eqref{wight} one can see that the 
response function \eqref{transition_rate} splits into the response function of
a detector coupled to a local massless scalar field, $F_0$, plus the response function of 
a local massive scalar field, $F_\mu$, 
integrated over $\mu$ weighted by $\rho(\mu^2)$.
We can therefore write \eqref{transition_rate} as 
\be
\mathcal{F}(\Omega, T)=F_0(\Omega, T)+\int_0^\infty d\mu^2\rho(\mu^2)F_\mu(\Omega, T).
\ee
Since the first term is common to both local and nonlocal theories, in what follows we will
study the relative difference in the detector's response, i.e.,
\be\label{relresp}
\Delta(l_n,\Omega,T):=
\frac{\mathcal{F}(\Omega,T)-F_{0}(\Omega,T)}{F_{0}(\Omega,T)}.
\ee

\begin{table}[]
\centering
%\caption{My caption}
\begin{tabular}{|c||c|c|c|m{10ex}|}
\hline
$\boldsymbol{\chi(t)}$ & $\boldsymbol{e^{-|t|}}$ & $\boldsymbol{\frac{\sin(t)}{t}}$ & $\boldsymbol{\frac{1}{t^2+1}}$  & $\;\,\boldsymbol{e^{-t^2}}$ \\
\hline
\hline
& & & & \\[-3mm]
$\boldsymbol{\Omega>0,\;\Omega T\gg 1}$ &  $\approx l_{n}^{2}/T^{2}$  & 0 & $\frac{l_{n}^{2}}{T^{2}}e^{-2\Omega T}$ & $ \frac{e^{-\frac{\Omega^{2}T^{2}}{2}}l_{n}^{2}}{\Omega^{4}T^{6}}$  \\[1mm]
\hline
& & & & \\[-3mm]
$\boldsymbol{|\Omega| T\ll 1}$ &  $\approx l_{n}^{2}/T^{2}$ & $l_{n}^{2}/T^{2}$  & $l_{n}^{2}/T^{2}$ & $\;\;\, l_{n}^{2}/T^{2}$ \\
\hline
& & & & \\[-3mm]
$\boldsymbol{\Omega<0,\;|\Omega| T\gg 1}$ & $Tl_{n}^{2}|\Omega|^{3}$ & $Tl_{n}^{2}|\Omega|^{3}$ & $ Tl_{n}^{2}|\Omega|^{3}$ & $\;\;\; Tl_{n}^{2}|\Omega|^{3}$\\
\hline
\end{tabular}
\caption{Detector's response $\mathcal{F}-F_{0}$ for various switching functions (taking a dimensionless argument $t=\tau/T$) and for both the exponential spectral function eq.\eqref{2disc} and the causal sets inspired spectral function eq.\eqref{discfunc2d}.}\label{table1}
\end{table}

It is a well known fact that in a local QFT an inertial detector in the ground state, switched on for an infinite time, $T\rightarrow\infty$, will not click 
because of Poincar\'e invariance. A straightforward calculation along the lines of~\cite{Birrell1984} 
shows that this is also true 
in the nonlocal theories studied in this paper. This should not come 
as a surprise given that such theories are also Poincar\'e invariant (and stable) by construction. 
We now ask what happens when the inertial detector is switched on for a 
finite time, $T$, which we implement by inserting non-trivial switching functions 
$\chi(\tau /T)$ in the Unruh-Dewitt interaction. Within this context, the most interesting case is that of spontaneous emission, i.e. when the detector starts out in an excited state, since in this case there can be differences between the behaviour of the detector coupled to local and nonlocal field theories even in the limit $T\rightarrow\infty$. Furthermore, spontaneous emission is a well-understood, experimentally accessible
phenomenon~\cite{Scully1997}.

We will assume that the nonlocality (length) scale is much smaller than any other length scale in the problem. In particular we assume that
%it is reasonable to think of nonlocality length scales below the nuclear scale~\cite{Sorkin:2007qi}, which are in any case well below any other scale in the detector dynamics. 
%These considerations impose the following hierarchy for the various scales in the problem: 
$|\Omega| l_{n}\ll 1,\;T/l_{n}\gg 1$,
where the first condition defines the ``low energy" condition, and the second ensures that the detector is switched on for a reasonable amount of time. 
\begin{figure*}
\includegraphics[width=0.31\textwidth]{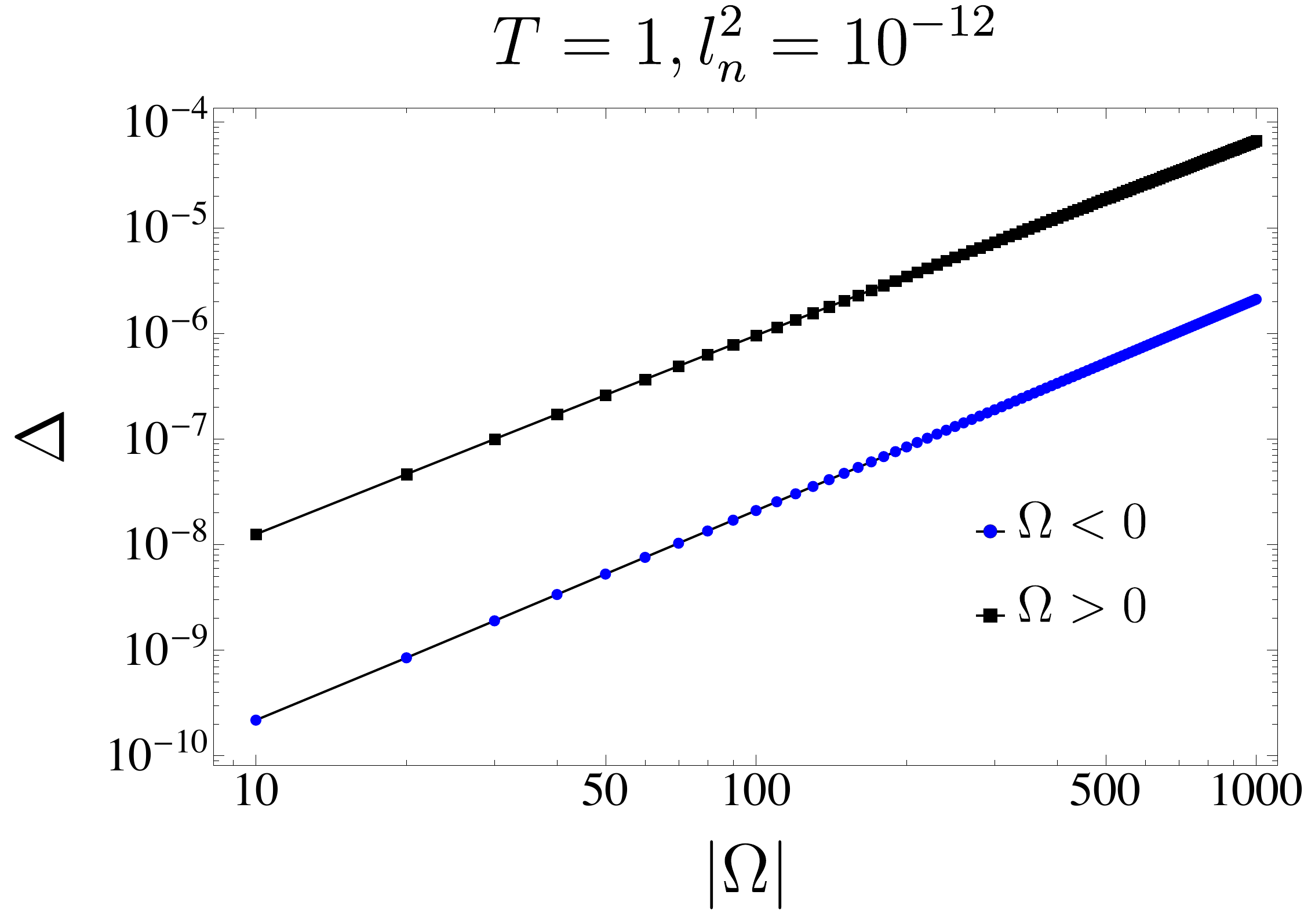}\quad
\includegraphics[width=0.32\textwidth]{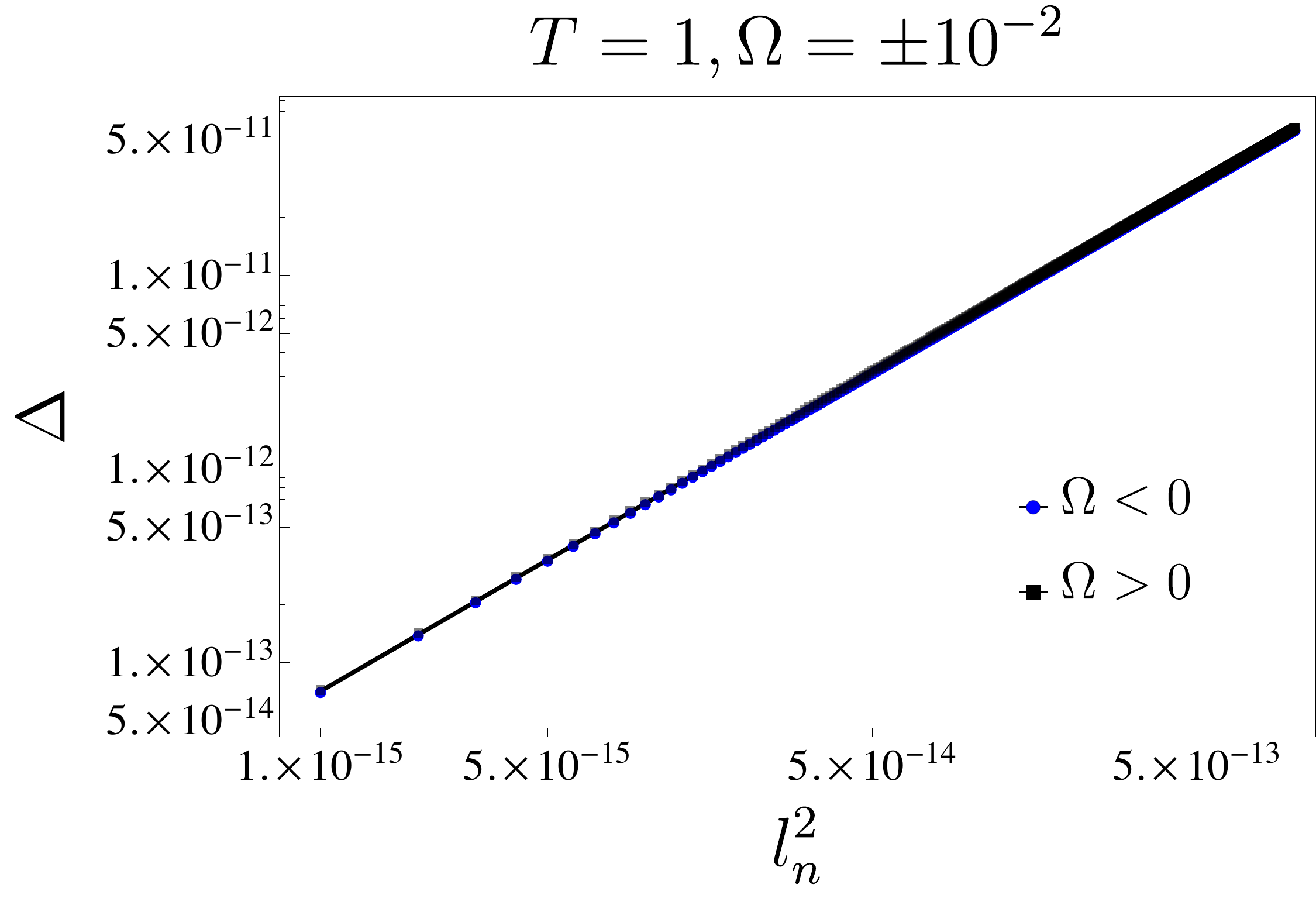}\quad
\includegraphics[width=0.33\textwidth]{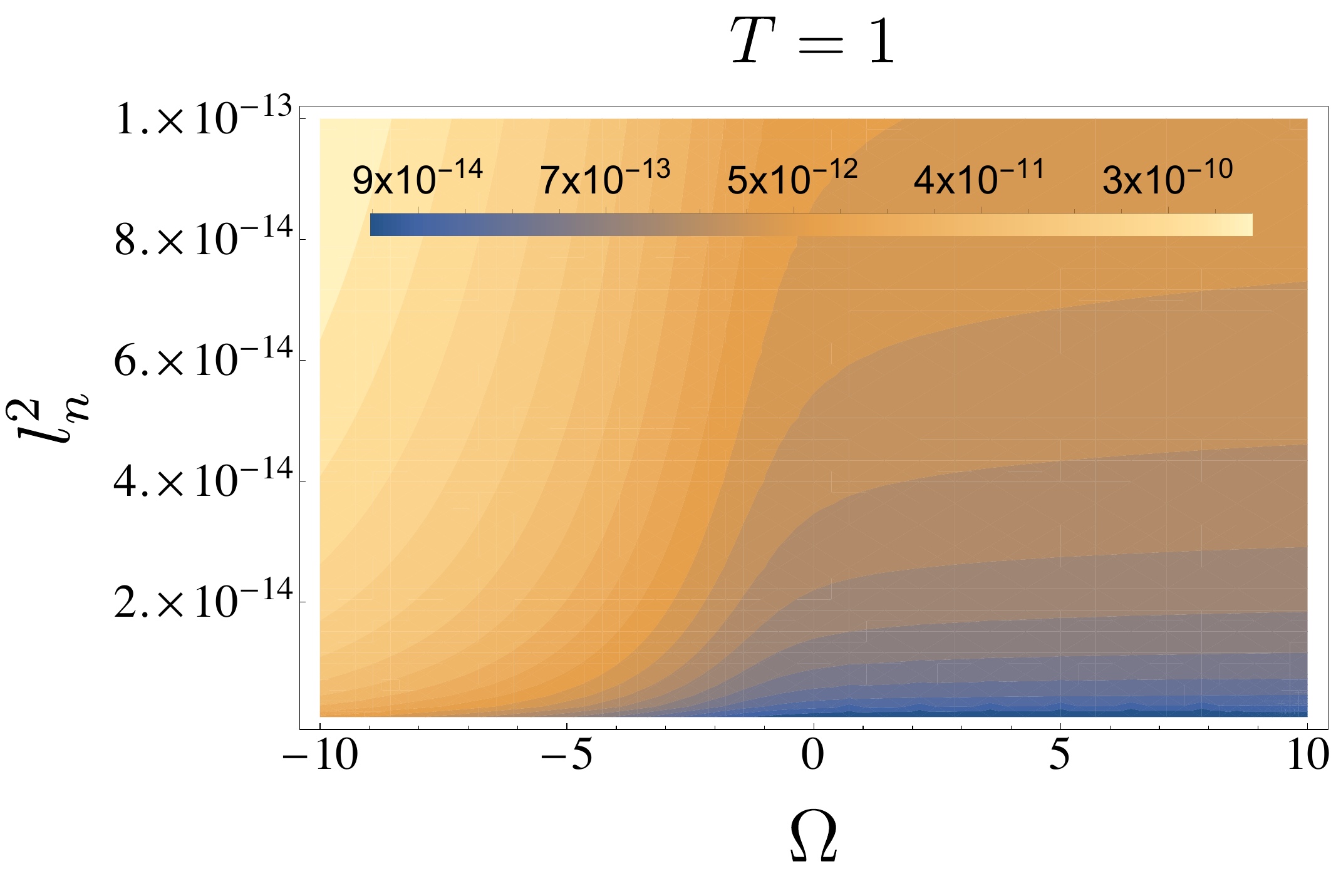}
  \caption{
  (Color online.)  Detector's relative response, $\Delta$ (eq.\eqref{relresp}) for the exponential switching function and spectral function eq.\eqref{discfunc2d}. From left to right we have: a) $|\Omega|T\gg 1$ for both positive (blue circles) and negative (black squares) $\Omega$, i.e. vacuum noise and spontaneous emission respectively; b) $|\Omega|T\ll 1$ for both positive (blue circles) and negative (black squares) $\Omega$. The two data sets overlap which is consistent with the behaviour reported in table \ref{table1} for  $|\Omega| T\ll1$; c) Logarithmic-scaled contour plot of $\Delta$. 
  %for both positive and negative $\Omega$. 
  Note from plot (a) that although the vacuum response ($\Omega>0$) has a larger relative difference compared to spontaneous emission ($\Omega<0$), measuring the latter is experimentally easier.
  }
  \label{FigQubit}
\end{figure*}
We first consider the behaviour of the detector's vacuum response and the spontaneous emission for short detector timescales, where we are able to perform an analytical analysis of the dependence of the results on the shape of the switching function.
Secondly we will analyze the more relevant and experimentally accessible case of spontaneous emission when the detector interacts with the field for long times compared to the detector's Heisenberg time $\Omega^{-1}$. In this experimentally accessible regime the detector's response is independent of the details of the switching function. We will show how a low energy detector can resolve nonlocality scales with a precision comparable to a high-energy particle collider experiment.

%both gaussian and exponential switching functions ({\bf DB: maybe more?}\textcolor{blue}{\textbf{AB:Yes}}),
%for which we find qualitatively similar outcomes. This hints at the fact that the qualitative
%behaviour of the response function in the nonlocal QFTs under consideration is independent
%of the specific form of the switching function of the detector.

\section{VACUUM RESPONSE AND SHORT TIME ($|\Omega|T\ll 1$) SPONTANEOUS EMISSION}%{\bf \textit{Vacuum response and short time ($|\Omega|T\ll 1$) spontaneous emission.-}} 
The behaviour of the relative response \eqref{relresp} can be readily analyzed for $|\Omega|T\ll 1$ regardless of the sign of $\Omega$. This regime corresponds to a  rapid switching of the detector. 
%It is possible to do a fully analytic study of $\Delta$ when $|\Omega|T\ll 1$.
On the other hand, a full analytical treatment in the case where  $\Omega>0$ (corresponding to studying the detector's spontaneous excitation probability due to the `vacuum noise' of the field)   and $\Omega T\gg 1$ has proven elusive. We summarize our findings in Table~\ref{table1}.

%Concretely, when $|\Omega|T\ll 1$, we can take the leading order approximation and replace $\Omega=0$.  As for the discontinuity function, we can use the zeroth order approximation for the discontinuity function $\rho=l_{n}^2$, as per the discussion at the end of the previous section. After some algebra, eq.\eqref{longeq} becomes
%\begin{align}
%\mathcal{F}_{nonlocal}-\mathcal{F}_{0}=\frac{4\pi}{3}\frac{l_{n}^2}{T^2}\int_{0}^{\infty}dk^0 (k^0)^3~|\tilde\chi(k^0)|^2.
%\end{align}
%Analogously, for the local, massless field --- obtained by replacing $\rho(\mu^2)$ with $\delta(\mu^2)$ --- we get
%\begin{equation}
%I_0=4\pi\int_0^{\infty}dx~x|\tilde \zeta(x)|^2
%\end{equation} 
%Of course, for this treatment to be valid  it is necessary that $
%\int_{0} ^{\infty}dx ~x^3~|\tilde\zeta(x)|^2<\infty.$
%The exponential switching function, considered in Tab.\ref{table1}, does not satisfy this condition. However, it is still possible to estimate the behaviour of the detector's response by introducing a cutoff in the integral over $\mu$. In such a case it turns out that the relative response at leading order is the same, i.e. 
%\begin{equation}\label{smallOmega}
%\Delta\propto  l_{n}^{2}(c T)^{-2},
%\end{equation}
%where we have reintroduced again the speed of light.

\section{SPONTANEOUS EMISSION}%{\bf \textit{Spontaneous emission.-}} 
Consider now the case in which $\Omega<0$, corresponding to the process of spontaneous emission. We are interested here in the regime characterized by $|\Omega|T\gg 1$, which corresponds to assuming that the detector is turned on for times much larger than the Heisenberg time of the atomic system. In this regime, we expect the detector's response to be largely independent of the specific form of the switching function.

Using the following dimensionless variables  \mbox{$
t=\tau/T$}, \mbox{$k=Tp$}, \mbox{$m=T\mu$},
and defining the Fourier transform of the switching function as $
\widetilde\chi(\omega)=\int dt~e^{-i\omega t}\chi(t)$,
one can show that 
\begin{align}\label{longeq}
\mathcal{F}-F_{0}=&\frac{1}{T^2}\int dm^2\rho(m^2/T^2)\\ \nonumber
&\times\int d^4k~\delta(k^2+m^2)|\widetilde\chi(k^0+\Omega T)|^2.
\end{align}
For switching functions whose Fourier transform decays asymptotically faster than polynomially (e.g. Gaussian, Lorentzian or sinc), and assuming that $|\Omega|T\gg 1$ and $|\Omega|l_n\ll 1$, we get the asymptotic result
\begin{equation}\label{emissionresponse}
\mathcal{F}-F_{0}\approx\frac{4\pi}{3}Tl_n^2|\Omega|^3\bigg(\big[1+\mathcal{O}(l^2\Omega^2)\big]\!\!\int_{-\infty}^{\infty}\!\!\!\!\text{d}x |\widetilde\chi(x)|^2\bigg).
\end{equation}
Performing a similar calculation in the local, massless case yields $F_0$. Finally we find that the relative response of eq.~\eqref{relresp}
goes like
\begin{equation}\label{emission}
\Delta(l_n,\Omega,T)\approx c^{-2}|\Omega|^{2}l_{n}^{2},
\end{equation}
where we have reintroduced the speed of light for dimensional reasons. This asymptotic expression should hold for any spectral function that is exponentially suppressed with $l_n$, such as \eqref{discfunc2d} and \eqref{2disc}. Although, rigorously speaking,  \eqref{emissionresponse}  was obtained for the exponential spectral function \eqref{2disc}, the asymptotic result \eqref{emission} is also confirmed by a numerical analysis with the spectral function in \eqref{discfunc2d} (see Tab. \ref{table1}). Finally, we note that the use of switching functions whose Fourier transform decays faster than polynomially is just a sufficient condition for \eqref{emission} to hold: We can see in Table \ref{table1} that \eqref{emission} also applies to all the switching modalities considered, including the exponential switching function, whose Fourier transform decays polynomially.

\section{DISCUSSION}%{\bf \textit{Discussion.-}} 
In all the  physically reasonable regimes studied for an inertial detector coupled for a finite time to a nonlocal field, we find that the nonlocal contribution to the detector's response is polynomial in the nonlocality scale, i.e. $\propto l_{n}^{2}$. The behaviour of the relative response (eq.~\eqref{relresp}) in different regimes is reported in Fig.~\ref{FigQubit}. This result is independent of the specific form of the switching function $\chi(t)$. The fact that nonlocal effects are not exponentially suppressed opens up interesting phenomenological windows.

In the case $|\Omega|T\ll 1$, we see from Table~\ref{table1} that the detector's response (and, indeed, also the relative response) is independent of the detector's gap, at leading order. As for the case of vacuum excitation with large $\Omega T$ we find that while the polynomial scaling with the nonlocality scale persists, the response is in general dependent on the details of the switching function. This is not surprising given that the a non-trivial dependence also occurs in the standard local, massless case.%Furthermore, note that the case $\Omega=0$ corresponds to the case of a detector with degenerate levels. %({\bf We can add some physical realization of such a regime.}).

In the case of spontaneous emission with $|\Omega|T\gg 1$, we see from eq. \eqref{emissionresponse} that the nonlocal contribution to the detector's response grows like $T\Omega^3$, a fact which can be used to amplify the signature of nonlocality in an experimental setting. Note that this regime is particularly interesting because spontaneous emission for times greater than the detector's Heisenberg time is an experimentally very well understood process~\cite{Scully1997} (indeed spontaneous emission is far easier to observe than vacuum noise). 
%The new scales introduced in the problem -- namely, the detector's gap and the finiteness of the interaction-- can be used to amplify the  signature of nonlocality. 
%tuning the detector parameters in an appropriate way. 

Substituting in some realistic numbers we can estimate the expected magnitude of the nonlocal signal. 
%and how difficult it would be to detect. 
Consider an experimental tolerance for the relative response  of $\Delta\sim 10^{-10}$. Such a tolerance implies that the experimenter has the ability to repeat the experiment of the order of billions of times and accumulate statistics in order to distinguish between the two probability distributions. Then using a frequency gap of the order of $10^{22}$Hz, corresponding to $\gamma$-ray transitions, we can cast a bound on $l_{n}\lesssim 10^{-19}$m. Note that this constraint is of the same order as present constraints on nonlocality coming from LHC data~\cite{Biswas:2014yia}. 

At first sight these numbers may seem experimentally far fetched, but recall that we are analyzing the process of spontaneous emission and we could have a large number of events. Let us analyze some realistic experimental testbeds in nuclear physics. Consider for example $^{20}_{11}$Na. This nuclear species has a half-life of $T_{1/2}\sim$500 ms and decays into  electromagnetically excited, highly unstable, $^{20}_{10}$Ne, which then spontaneously decays to its ground state emitting $\sim 11$ MeV gamma radiation \cite{Web,Ingalls:1976zz}. Suppose now that one has $\sim 20$ grams of $^{20}_{11}$Na ($\sim N_\text{A}\approx 6\times 10^{23}$ atoms), then according to the radioactive decay laws the number of gamma emission events in time $\tau$ is given by
\begin{equation}
N_\gamma\big(\tau)=N_\text{A}(1-e^{-\frac{\tau}{T_{1/2}}\frac{1}{\ln2}}\big).
\end{equation}
But in a time of $\tau\sim 10$s, $N_\gamma\simeq N_\text{A}\sim 10^{23}\gg\Delta^{-1}$.
%We can consider the fact that 
Assuming that gamma ray detection is  not 100\% efficient (which it is not), and in particular assuming a very conservative 0.1 \% experimental detection efficiency (at least one order of magnitude more conservative than realistic estimates \cite{SINGH1971475}),
%Even in the extremely conservative assumption that we cannot see all the gamma ray emitted by the nuclear 'detectors' and that the experimental detection efficiency of the gamma rays themselves were below 0.1 \%, 
there are still orders of magnitude more detection events than $\Delta^{-1}$. 
In other words a low energy  nuclear physics experiment ($\sim$10 MeV scale) would already yield a higher resolution than the LHC experiments. In theory, following the reasoning above, if we assume that we have 200 grams $^{20}_{11}$Na (i.e., we have $\sim 10 N_A$ of the nuclear species) a very conservative estimate for this number of emission events yields that the detectable relative response would be of order $\Delta\sim 10^{-23}$, which in turn implies that the experiment could detect nonlocality scales of $l_n\lesssim10^{-25}$ m, many orders of magnitude better than the resolution of the LHC.
Furthermore, there are more than a dozen different nuclear species that provide a reliable source of spontaneous emission of gamma rays \cite{Web}, so the use of $^{20}_{11}$Na provides just one possible example.

%It should be noted that, due to the similarity of Eq.~\eqref{wight} with the standard K\"all\'en-Lehmann representation for interacting theories~\cite{itzykson2006quantum}, 
Due to the similarity of Eq.~\eqref{wight} with the standard K\"all\'en-Lehmann representation for interacting theories~\cite{itzykson2006quantum}, one may wonder whether it is possible to discern the nonlocal contribution to spontaneous emission from the similar effect that would arise through interaction with a secondary massive field. In fact, 
%interacting with the massless probed one. 
one can show that such a contribution, in the case of long time spontaneous emission, vanishes unless the massive field's mass $2m<|\Omega|$. Therefore, for EM nuclear decay,
%is below the detector's gap. 
%In the case of spontaneous emission in nuclei, 
%it would suffice to 
considering $|\Omega|<2 m_e\sim 1$ MeV
%, the mass of the electron, 
%(the lightest charged field). 
would suffice to guarantee that the only non-trivial
%massive 
contribution to~\eqref{transition_rate} %\textcolor{red}{[MS: for $|\Omega|<m_e$ the only nontrivial contribution to the transition rate eq. (7) or eq. (8) and not eq. (4) comes from the nonlocality (because there is always contribution to the Wightman function, but it doesn't contribute to the transition rate]}
comes from $l_n$. 
%In the case of an electromagnetic nueclear decay, one can achieve this by considering $|\Omega|\lesssim$ 1MeV. 
%Furthermore, 
Doing so would worsen the bound on $l_n$ discussed above by 
one order of magnitude -- which is still better than the LHC bound -- but it would also
greatly increase the number of experimentally viable nuclear species.
%at the expense of worsening the bound on $l_n$ discussed above by one order of magnitude, still much more ambitious than the LHC bound.
%
%Experimentally this means considering gaps of 1 MeV, 
%instead of 10 MeV. 
%If we do so, we would lose 
%which does worsen the bound on $l_n$ by 
%one order of magnitude, 
%in the upper bound to the nonlocality scale,
%but in exchange, 
%but also greatly increases the number of experimentally viable nuclear species.
%there are many more nuclear species that can be used to make the experiment more feasible to implement. 
Furthermore, contributions from local %interactions with 
massive fields can in principle always be %taken into account 
accounted for \textit{a priori} and subtracted when defining $\Delta$ (see Eq.~\eqref{relresp}). 
%since, from the previous argument, only known massive species will contribute to the detector response.

%One can even think of other possible implementations in even lower energy scales, for example soft x-ray coherent spontaneous emission \cite{} taht woul.

%({\bf AB: is the numerical prefactor in $\Delta$ order one? or does it depends on the switching function? in the second case can we make better constraints? Furthermore, given that we are studying the relative response, increasing the number of detectors can really amplify the signal?})   

%{\bf DB: This section will very much depend on what discontinuity functions and switching functions we consider.
%Ideally I think it would be nice if we could provide partial evidence towards a generic $l^2$ dependence by
%showing that many choices for the switching function lead to the same qualitative features. This could be strengthened
%by a (semi) analytic analysis (which we already have) of the exponential discontinuity functions if we are able to 
%convincingly argue that qualitative properties of the causet functions are captures by this simpler case. In any case
%the punch line of our results is the dependence $l^2\Omega^2$ in the limit $\Omega T\gg 1$ and $l^2/T^2$ in the 
%limit $\Omega T\ll1$.}
%\textcolor{blue}{\textbf{AB:Given the new results concerning other switching functions we can argue for the generality of the power-law dependence on the nonlocality scale. Moreover, with the new results it seems apparent that the exponential spectral function captures preatty well the qualitative features of the causal set one.}}

\section{CONCLUSION}%{\bf \textit{Conclusions.-}}
%We have studied the response of an inertial particle detector coupled to a nonlocal field theory. 
%{\red Although the Unruh--DeWitt model of a particle detector is an idealized system, it still captures the fundamental features of matter-light interactions, and it has  already been proven to be low energy systems sensible to high-energy Lorentz invariant violations~\cite{ViqarJorma}.}
We have studied the low energy response of particle detectors coupled to a Lorentz Invariant nonlocal QFTs characterized by a nonanalytic functions of $\Box$, a kind of nonlocality that finds its roots in models of LI discrete spacetimes~\cite{Sorkin:2007qi,Aslanbeigi:2014zva,Belenchia:2014fda}.
%, and is phenomenologically  attractive 
%for possible models of quantum gravity due to the otherwise strict constraints on violations of Lorentz symmetry~\cite{Liberati:2013xla}. 
For the cases considered (eqns. \eqref{discfunc2d} and \eqref{2disc}), we gave both numerical and analytical evidence that the detector's relative response depends quadratically on the nonlocality scale, and argued that this result should hold for any exponentially suppressed spectral function $\rho$.

%We find that the detector can become highly sensitive to the nonlocality scale of the theory. In particular, we have found corrections to spontaneous emission of radiation due to the presence of a nonlocality scale. 

We exploited this fact to show that experimentally feasible setups -- involving detectors with energy gaps of the order of MeVs (e.g. gamma emission following the $\beta$ decay of  $^{20}_{11}$Na) -- can potentially  probe nonlocality scales of the order of $l_n\lesssim 10^{-25}$ m, six orders of magnitude better than a TeV-scale experiment at the LHC~\cite{Biswas:2014yia}. This paves the way for low energy experimental tests of high-energy theories and models of quantum gravity.

\section*{ACNOWLEDGEMENTS}%{\bf \textit{Acknowledgements.-}} 
AB and DMTB  would like to acknowledge financial support from the John Templeton Foundation (JTF), grant  No. 51876. E.M-M effusively thanks the hospitality of Stefano Liberati during the time Stefano hosted him at SISSA and for helpful discussions. E.M-M was partially funded by the NSERC Discovery grant programme. E. M-M would like to thank Achim Kempf for his helpful insights. MS would like to thank Ted Jacobson for helpful discussions. The authors would also like to thank Jorma Louko for his useful comments on this work. MS was supported in part by Perimeter Institute for Theoretical Physics. Research at Perimeter Institute is supported by the Government of Canada through Industry Canada and by the Province of Ontario through the Ministry of Research and Innovation.

\bibliography{references}

\begin{thebibliography}{33}%
\makeatletter
\providecommand \@ifxundefined [1]{%
 \@ifx{#1\undefined}
}%
\providecommand \@ifnum [1]{%
 \ifnum #1\expandafter \@firstoftwo
 \else \expandafter \@secondoftwo
 \fi
}%
\providecommand \@ifx [1]{%
 \ifx #1\expandafter \@firstoftwo
 \else \expandafter \@secondoftwo
 \fi
}%
\providecommand \natexlab [1]{#1}%
\providecommand \enquote  [1]{``#1''}%
\providecommand \bibnamefont  [1]{#1}%
\providecommand \bibfnamefont [1]{#1}%
\providecommand \citenamefont [1]{#1}%
\providecommand \href@noop [0]{\@secondoftwo}%
\providecommand \href [0]{\begingroup \@sanitize@url \@href}%
\providecommand \@href[1]{\@@startlink{#1}\@@href}%
\providecommand \@@href[1]{\endgroup#1\@@endlink}%
\providecommand \@sanitize@url [0]{\catcode `\\12\catcode `\$12\catcode
  `\&12\catcode `\#12\catcode `\^12\catcode `\_12\catcode `\%12\relax}%
\providecommand \@@startlink[1]{}%
\providecommand \@@endlink[0]{}%
\providecommand \url  [0]{\begingroup\@sanitize@url \@url }%
\providecommand \@url [1]{\endgroup\@href {#1}{\urlprefix }}%
\providecommand \urlprefix  [0]{URL }%
\providecommand \Eprint [0]{\href }%
\providecommand \doibase [0]{http://dx.doi.org/}%
\providecommand \selectlanguage [0]{\@gobble}%
\providecommand \bibinfo  [0]{\@secondoftwo}%
\providecommand \bibfield  [0]{\@secondoftwo}%
\providecommand \translation [1]{[#1]}%
\providecommand \BibitemOpen [0]{}%
\providecommand \bibitemStop [0]{}%
\providecommand \bibitemNoStop [0]{.\EOS\space}%
\providecommand \EOS [0]{\spacefactor3000\relax}%
\providecommand \BibitemShut  [1]{\csname bibitem#1\endcsname}%
\let\auto@bib@innerbib\@empty
%</preamble>
\bibitem [{\citenamefont {Pais}\ and\ \citenamefont
  {Uhlenbeck}(1950)}]{Pais:1950za}%
  \BibitemOpen
  \bibfield  {author} {\bibinfo {author} {\bibfnamefont {A.}~\bibnamefont
  {Pais}}\ and\ \bibinfo {author} {\bibfnamefont {G.}~\bibnamefont
  {Uhlenbeck}},\ }\href {\doibase 10.1103/PhysRev.79.145} {\bibfield  {journal}
  {\bibinfo  {journal} {Phys.Rev.}\ }\textbf {\bibinfo {volume} {79}},\
  \bibinfo {pages} {145} (\bibinfo {year} {1950})}\BibitemShut {NoStop}%
%%CITATION = PHRVA,79,145;%%
\bibitem [{\citenamefont {Sorkin}(2007)}]{Sorkin:2007qi}%
  \BibitemOpen
  \bibfield  {author} {\bibinfo {author} {\bibfnamefont {R.~D.}\ \bibnamefont
  {Sorkin}},\ }\href@noop {} {\  (\bibinfo {year} {2007})},\ \Eprint
  {http://arxiv.org/abs/gr-qc/0703099} {arXiv:gr-qc/0703099 [GR-QC]}
  \BibitemShut {NoStop}%
%%CITATION = GR-QC/0703099;%%
\bibitem [{\citenamefont {Moeller}\ and\ \citenamefont
  {Zwiebach}(2002)}]{Moeller:2002vx}%
  \BibitemOpen
  \bibfield  {author} {\bibinfo {author} {\bibfnamefont {N.}~\bibnamefont
  {Moeller}}\ and\ \bibinfo {author} {\bibfnamefont {B.}~\bibnamefont
  {Zwiebach}},\ }\href {\doibase 10.1088/1126-6708/2002/10/034} {\bibfield
  {journal} {\bibinfo  {journal} {J. High Energy Phys.}\ }\textbf {\bibinfo
  {volume} {10}},\ \bibinfo {pages} {034} (\bibinfo {year} {2002})}\BibitemShut
  {NoStop}%
%%CITATION = HEP-TH/0207107;%%
\bibitem [{\citenamefont {Gambini}\ and\ \citenamefont
  {Pullin}(2014)}]{Gambini:2014kba}%
  \BibitemOpen
  \bibfield  {author} {\bibinfo {author} {\bibfnamefont {R.}~\bibnamefont
  {Gambini}}\ and\ \bibinfo {author} {\bibfnamefont {J.}~\bibnamefont
  {Pullin}},\ }\href@noop {} {\bibfield  {journal} {\bibinfo  {journal} {Int.
  J. Mod. Phys. D}\ }\textbf {\bibinfo {volume} {23}},\ \bibinfo {pages}
  {1442023} (\bibinfo {year} {2014})}\BibitemShut {NoStop}%
\bibitem [{\citenamefont {Modesto}(2011)}]{Modesto:2011aa}%
  \BibitemOpen
  \bibfield  {author} {\bibinfo {author} {\bibfnamefont {L.}~\bibnamefont
  {Modesto}},\ }\href {http://arxiv.org/abs/1107.2403} {\  (\bibinfo {year}
  {2011})},\ \Eprint {http://arxiv.org/abs/ArXiv:1107.2403} {ArXiv:1107.2403}
  \BibitemShut {NoStop}%
\bibitem [{\citenamefont {Efimov}(1967)}]{Efimov1967}%
  \BibitemOpen
  \bibfield  {author} {\bibinfo {author} {\bibfnamefont {G.~V.}\ \bibnamefont
  {Efimov}},\ }\href {\doibase 10.1007/BF01646357} {\bibfield  {journal}
  {\bibinfo  {journal} {Communications in Mathematical Physics}\ }\textbf
  {\bibinfo {volume} {5}},\ \bibinfo {pages} {42} (\bibinfo {year}
  {1967})}\BibitemShut {NoStop}%
\bibitem [{\citenamefont {Aslanbeigi}\ \emph {et~al.}(2014)\citenamefont
  {Aslanbeigi}, \citenamefont {Saravani},\ and\ \citenamefont
  {Sorkin}}]{Aslanbeigi:2014zva}%
  \BibitemOpen
  \bibfield  {author} {\bibinfo {author} {\bibfnamefont {S.}~\bibnamefont
  {Aslanbeigi}}, \bibinfo {author} {\bibfnamefont {M.}~\bibnamefont
  {Saravani}}, \ and\ \bibinfo {author} {\bibfnamefont {R.~D.}\ \bibnamefont
  {Sorkin}},\ }\href {\doibase 10.1007/JHEP06(2014)024} {\bibfield  {journal}
  {\bibinfo  {journal} {J. High Energy Phys.}\ }\textbf {\bibinfo {volume}
  {06}},\ \bibinfo {pages} {024} (\bibinfo {year} {2014})}\BibitemShut
  {NoStop}%
%%CITATION = ARXIV:1403.1622;%%
\bibitem [{\citenamefont {Weinberg}(2005)}]{Weinberg:2005}%
  \BibitemOpen
  \bibfield  {author} {\bibinfo {author} {\bibfnamefont {S.}~\bibnamefont
  {Weinberg}},\ }\href@noop {} {\emph {\bibinfo {title} {The Quantum Theory of
  Fields, Volume 1: Foundations}}}\ (\bibinfo  {publisher} {Cambridge
  University Press},\ \bibinfo {year} {2005})\BibitemShut {NoStop}%
\bibitem [{\citenamefont {Gomis}\ and\ \citenamefont
  {Mehen}(2000)}]{Gomis:2000aa}%
  \BibitemOpen
  \bibfield  {author} {\bibinfo {author} {\bibfnamefont {J.}~\bibnamefont
  {Gomis}}\ and\ \bibinfo {author} {\bibfnamefont {T.}~\bibnamefont {Mehen}},\
  }\href {http://arxiv.org/abs/hep-th/0005129} {\bibfield  {journal} {\bibinfo
  {journal} {Nucl. Phys. B}\ }\textbf {\bibinfo {volume} {591}},\ \bibinfo
  {pages} {265} (\bibinfo {year} {2000})},\ \Eprint
  {http://arxiv.org/abs/hep-th/0005129} {hep-th/0005129} \BibitemShut {NoStop}%
\bibitem [{\citenamefont {Alvarez-Gaume}\ \emph {et~al.}(2001)\citenamefont
  {Alvarez-Gaume}, \citenamefont {Barbon},\ and\ \citenamefont
  {Zwicky}}]{AlvarezGaume:2001ka}%
  \BibitemOpen
  \bibfield  {author} {\bibinfo {author} {\bibfnamefont {L.}~\bibnamefont
  {Alvarez-Gaume}}, \bibinfo {author} {\bibfnamefont {J.~L.~F.}\ \bibnamefont
  {Barbon}}, \ and\ \bibinfo {author} {\bibfnamefont {R.}~\bibnamefont
  {Zwicky}},\ }\href@noop {} {\bibfield  {journal} {\bibinfo  {journal} {JHEP}\
  }\textbf {\bibinfo {volume} {05}},\ \bibinfo {pages} {057} (\bibinfo {year}
  {2001})}\BibitemShut {NoStop}%
\bibitem [{\citenamefont {Barci}\ and\ \citenamefont
  {Oxman}(1997)}]{Barci:1997xy}%
  \BibitemOpen
  \bibfield  {author} {\bibinfo {author} {\bibfnamefont {D.~G.}\ \bibnamefont
  {Barci}}\ and\ \bibinfo {author} {\bibfnamefont {L.~E.}\ \bibnamefont
  {Oxman}},\ }\href@noop {} {\bibfield  {journal} {\bibinfo  {journal} {Mod.
  Phys. Lett. A}\ }\textbf {\bibinfo {volume} {12}},\ \bibinfo {pages} {493}
  (\bibinfo {year} {1997})}\BibitemShut {NoStop}%
\bibitem [{\citenamefont {Biswas}\ and\ \citenamefont
  {Okada}(2015)}]{Biswas:2014yia}%
  \BibitemOpen
  \bibfield  {author} {\bibinfo {author} {\bibfnamefont {T.}~\bibnamefont
  {Biswas}}\ and\ \bibinfo {author} {\bibfnamefont {N.}~\bibnamefont {Okada}},\
  }\href {\doibase 10.1016/j.nuclphysb.2015.06.023} {\bibfield  {journal}
  {\bibinfo  {journal} {Nucl. Phys. B}\ }\textbf {\bibinfo {volume} {898}},\
  \bibinfo {pages} {113} (\bibinfo {year} {2015})}\BibitemShut {NoStop}%
%%CITATION = ARXIV:1407.3331;%%
\bibitem [{\citenamefont {Belenchia}\ \emph {et~al.}(2016)\citenamefont
  {Belenchia}, \citenamefont {Benincasa}, \citenamefont {Liberati},
  \citenamefont {Marin}, \citenamefont {Marino},\ and\ \citenamefont
  {Ortolan}}]{Belenchia:2015ake}%
  \BibitemOpen
  \bibfield  {author} {\bibinfo {author} {\bibfnamefont {A.}~\bibnamefont
  {Belenchia}}, \bibinfo {author} {\bibfnamefont {D.~M.~T.}\ \bibnamefont
  {Benincasa}}, \bibinfo {author} {\bibfnamefont {S.}~\bibnamefont {Liberati}},
  \bibinfo {author} {\bibfnamefont {F.}~\bibnamefont {Marin}}, \bibinfo
  {author} {\bibfnamefont {F.}~\bibnamefont {Marino}}, \ and\ \bibinfo {author}
  {\bibfnamefont {A.}~\bibnamefont {Ortolan}},\ }\href {\doibase
  10.1103/PhysRevLett.116.161303} {\bibfield  {journal} {\bibinfo  {journal}
  {Phys. Rev. Lett.}\ }\textbf {\bibinfo {volume} {116}},\ \bibinfo {pages}
  {161303} (\bibinfo {year} {2016})}\BibitemShut {NoStop}%
%%CITATION = ARXIV:1512.02083;%%
\bibitem [{\citenamefont {Saravani}\ and\ \citenamefont
  {Aslanbeigi}(2015)}]{Saravani:2015rva}%
  \BibitemOpen
  \bibfield  {author} {\bibinfo {author} {\bibfnamefont {M.}~\bibnamefont
  {Saravani}}\ and\ \bibinfo {author} {\bibfnamefont {S.}~\bibnamefont
  {Aslanbeigi}},\ }\href {\doibase 10.1103/PhysRevD.92.103504} {\bibfield
  {journal} {\bibinfo  {journal} {Phys. Rev. D}\ }\textbf {\bibinfo {volume}
  {92}},\ \bibinfo {pages} {103504} (\bibinfo {year} {2015})}\BibitemShut
  {NoStop}%
%%CITATION = ARXIV:1502.01655;%%
\bibitem [{\citenamefont {Barnaby}\ and\ \citenamefont
  {Cline}(2008)}]{Barnaby:2008fk}%
  \BibitemOpen
  \bibfield  {author} {\bibinfo {author} {\bibfnamefont {N.}~\bibnamefont
  {Barnaby}}\ and\ \bibinfo {author} {\bibfnamefont {J.~M.}\ \bibnamefont
  {Cline}},\ }\href {\doibase 10.1088/1475-7516/2008/06/030} {\bibfield
  {journal} {\bibinfo  {journal} {JCAP}\ }\textbf {\bibinfo {volume} {0806}},\
  \bibinfo {pages} {030} (\bibinfo {year} {2008})},\ \Eprint
  {http://arxiv.org/abs/0802.3218} {arXiv:0802.3218 [hep-th]} \BibitemShut
  {NoStop}%
%%CITATION = ARXIV:0802.3218;%%
\bibitem [{\citenamefont {Kajuri}(2016)}]{Polymer}%
  \BibitemOpen
  \bibfield  {author} {\bibinfo {author} {\bibfnamefont {N.}~\bibnamefont
  {Kajuri}},\ }\href {http://stacks.iop.org/0264-9381/33/i=5/a=055007}
  {\bibfield  {journal} {\bibinfo  {journal} {Classical and Quantum Gravity}\
  }\textbf {\bibinfo {volume} {33}},\ \bibinfo {pages} {055007} (\bibinfo
  {year} {2016})}\BibitemShut {NoStop}%
\bibitem [{\citenamefont {Husain}\ and\ \citenamefont
  {Louko}(2016)}]{ViqarJorma}%
  \BibitemOpen
  \bibfield  {author} {\bibinfo {author} {\bibfnamefont {V.}~\bibnamefont
  {Husain}}\ and\ \bibinfo {author} {\bibfnamefont {J.}~\bibnamefont {Louko}},\
  }\href {\doibase 10.1103/PhysRevLett.116.061301} {\bibfield  {journal}
  {\bibinfo  {journal} {Phys. Rev. Lett.}\ }\textbf {\bibinfo {volume} {116}},\
  \bibinfo {pages} {061301} (\bibinfo {year} {2016})}\BibitemShut {NoStop}%
\bibitem [{\citenamefont {Mart\'{i}n-Mart\'{i}nez}\ \emph
  {et~al.}(2013)\citenamefont {Mart\'{i}n-Mart\'{i}nez}, \citenamefont
  {Montero},\ and\ \citenamefont {del Rey}}]{Martin-Martinez2013}%
  \BibitemOpen
  \bibfield  {author} {\bibinfo {author} {\bibfnamefont {E.}~\bibnamefont
  {Mart\'{i}n-Mart\'{i}nez}}, \bibinfo {author} {\bibfnamefont
  {M.}~\bibnamefont {Montero}}, \ and\ \bibinfo {author} {\bibfnamefont
  {M.}~\bibnamefont {del Rey}},\ }\href {\doibase 10.1103/PhysRevD.87.064038}
  {\bibfield  {journal} {\bibinfo  {journal} {Phys. Rev. D}\ }\textbf {\bibinfo
  {volume} {87}},\ \bibinfo {pages} {064038} (\bibinfo {year}
  {2013})}\BibitemShut {NoStop}%
\bibitem [{\citenamefont {Alhambra}\ \emph {et~al.}(2014)\citenamefont
  {Alhambra}, \citenamefont {Kempf},\ and\ \citenamefont
  {Mart\'in-Mart\'inez}}]{Alvaro}%
  \BibitemOpen
  \bibfield  {author} {\bibinfo {author} {\bibfnamefont {A.~M.}\ \bibnamefont
  {Alhambra}}, \bibinfo {author} {\bibfnamefont {A.}~\bibnamefont {Kempf}}, \
  and\ \bibinfo {author} {\bibfnamefont {E.}~\bibnamefont
  {Mart\'in-Mart\'inez}},\ }\href {\doibase 10.1103/PhysRevA.89.033835}
  {\bibfield  {journal} {\bibinfo  {journal} {Phys. Rev. A}\ }\textbf {\bibinfo
  {volume} {89}},\ \bibinfo {pages} {033835} (\bibinfo {year}
  {2014})}\BibitemShut {NoStop}%
\bibitem [{\citenamefont {Liberati}(2013)}]{Liberati:2013xla}%
  \BibitemOpen
  \bibfield  {author} {\bibinfo {author} {\bibfnamefont {S.}~\bibnamefont
  {Liberati}},\ }\href {\doibase 10.1088/0264-9381/30/13/133001} {\bibfield
  {journal} {\bibinfo  {journal} {Class. Quant. Grav.}\ }\textbf {\bibinfo
  {volume} {30}},\ \bibinfo {pages} {133001} (\bibinfo {year} {2013})},\
  \Eprint {http://arxiv.org/abs/1304.5795} {1304.5795} \BibitemShut {NoStop}%
%%CITATION = ARXIV:1304.5795;%%
\bibitem [{\citenamefont {Mattingly}(2005)}]{Mattingly:2005re}%
  \BibitemOpen
  \bibfield  {author} {\bibinfo {author} {\bibfnamefont {D.}~\bibnamefont
  {Mattingly}},\ }\href {\doibase 10.12942/lrr-2005-5} {\bibfield  {journal}
  {\bibinfo  {journal} {Living Rev. Rel.}\ }\textbf {\bibinfo {volume} {8}},\
  \bibinfo {pages} {5} (\bibinfo {year} {2005})}\BibitemShut {NoStop}%
%%CITATION = GR-QC/0502097;%%
\bibitem [{\citenamefont {Dom{\'\i}nguez}\ and\ \citenamefont
  {Trione}(1979)}]{dominguez1979laplace}%
  \BibitemOpen
  \bibfield  {author} {\bibinfo {author} {\bibfnamefont {A.~G.}\ \bibnamefont
  {Dom{\'\i}nguez}}\ and\ \bibinfo {author} {\bibfnamefont {S.~E.}\
  \bibnamefont {Trione}},\ }\href@noop {} {\bibfield  {journal} {\bibinfo
  {journal} {Advances in Mathematics}\ }\textbf {\bibinfo {volume} {31}},\
  \bibinfo {pages} {51} (\bibinfo {year} {1979})}\BibitemShut {NoStop}%
\bibitem [{\citenamefont {Benincasa}\ and\ \citenamefont
  {Dowker}(2010)}]{Benincasa:2010aa}%
  \BibitemOpen
  \bibfield  {author} {\bibinfo {author} {\bibfnamefont {D.~M.~T.}\
  \bibnamefont {Benincasa}}\ and\ \bibinfo {author} {\bibfnamefont
  {F.}~\bibnamefont {Dowker}},\ }\href {http://arxiv.org/abs/1001.2725}
  {\bibfield  {journal} {\bibinfo  {journal} {Phys. Rev. Lett.}\ }\textbf
  {\bibinfo {volume} {104}},\ \bibinfo {pages} {181301} (\bibinfo {year}
  {2010})},\ \Eprint {http://arxiv.org/abs/1001.2725} {1001.2725} \BibitemShut
  {NoStop}%
\bibitem [{\citenamefont {Dowker}\ and\ \citenamefont
  {Glaser}(2013)}]{Dowker:2013vl}%
  \BibitemOpen
  \bibfield  {author} {\bibinfo {author} {\bibfnamefont {F.}~\bibnamefont
  {Dowker}}\ and\ \bibinfo {author} {\bibfnamefont {L.}~\bibnamefont
  {Glaser}},\ }\href@noop {} {\bibfield  {journal} {\bibinfo  {journal} {Class.
  Quantum Grav.}\ }\textbf {\bibinfo {volume} {30}},\ \bibinfo {pages} {195016}
  (\bibinfo {year} {2013})}\BibitemShut {NoStop}%
\bibitem [{\citenamefont {Belenchia}\ \emph {et~al.}(2015)\citenamefont
  {Belenchia}, \citenamefont {Benincasa},\ and\ \citenamefont
  {Liberati}}]{Belenchia:2014fda}%
  \BibitemOpen
  \bibfield  {author} {\bibinfo {author} {\bibfnamefont {A.}~\bibnamefont
  {Belenchia}}, \bibinfo {author} {\bibfnamefont {D.~M.~T.}\ \bibnamefont
  {Benincasa}}, \ and\ \bibinfo {author} {\bibfnamefont {S.}~\bibnamefont
  {Liberati}},\ }\href {\doibase 10.1007/JHEP03(2015)036} {\bibfield  {journal}
  {\bibinfo  {journal} {J. High Energy Phys.}\ }\textbf {\bibinfo {volume}
  {03}},\ \bibinfo {pages} {036} (\bibinfo {year} {2015})}\BibitemShut
  {NoStop}%
%%CITATION = ARXIV:1411.6513;%%
\bibitem [{\citenamefont {Saravani}\ and\ \citenamefont
  {Afshordi}(2016)}]{Saravani:2016enc}%
  \BibitemOpen
  \bibfield  {author} {\bibinfo {author} {\bibfnamefont {M.}~\bibnamefont
  {Saravani}}\ and\ \bibinfo {author} {\bibfnamefont {N.}~\bibnamefont
  {Afshordi}},\ }\href@noop {} {\  (\bibinfo {year} {2016})},\ \Eprint
  {http://arxiv.org/abs/1604.02448} {arXiv:1604.02448 [gr-qc]} \BibitemShut
  {NoStop}%
%%CITATION = ARXIV:1604.02448;%%
\bibitem [{\citenamefont {Birrell}\ and\ \citenamefont
  {Davies}(1984)}]{Birrell1984}%
  \BibitemOpen
  \bibfield  {author} {\bibinfo {author} {\bibfnamefont {N.}~\bibnamefont
  {Birrell}}\ and\ \bibinfo {author} {\bibfnamefont {P.}~\bibnamefont
  {Davies}},\ }\href@noop {} {\emph {\bibinfo {title} {Quantum fields in curved
  space}}}\ (\bibinfo  {publisher} {Cambridge university press},\ \bibinfo
  {year} {1984})\BibitemShut {NoStop}%
\bibitem [{\citenamefont {Sriramkumar}\ and\ \citenamefont
  {Padmanabhan}(1996)}]{Sriramkumar:1994pb}%
  \BibitemOpen
  \bibfield  {author} {\bibinfo {author} {\bibfnamefont {L.}~\bibnamefont
  {Sriramkumar}}\ and\ \bibinfo {author} {\bibfnamefont {T.}~\bibnamefont
  {Padmanabhan}},\ }\href {\doibase 10.1088/0264-9381/13/8/005} {\bibfield
  {journal} {\bibinfo  {journal} {Class. Quant. Grav.}\ }\textbf {\bibinfo
  {volume} {13}},\ \bibinfo {pages} {2061} (\bibinfo {year}
  {1996})}\BibitemShut {NoStop}%
%%CITATION = GR-QC/9408037;%%
\bibitem [{\citenamefont {Scully}\ and\ \citenamefont
  {Zubairy}(1997)}]{Scully1997}%
  \BibitemOpen
  \bibfield  {author} {\bibinfo {author} {\bibfnamefont {M.}~\bibnamefont
  {Scully}}\ and\ \bibinfo {author} {\bibfnamefont {M.}~\bibnamefont
  {Zubairy}},\ }\href {http://books.google.ca/books?id=UtdfQgAACAAJ} {\emph
  {\bibinfo {title} {Quantum Optics}}}\ (\bibinfo  {publisher} {Cambridge
  University Press},\ \bibinfo {year} {1997})\BibitemShut {NoStop}%
\bibitem [{\citenamefont {{International Atomic Energy Agency, International
  Nuclear Structure and Decay Data Network.
  https://www-nds.iaea.org/}}()}]{Web}%
  \BibitemOpen
  \bibfield  {author} {\bibinfo {author} {\bibnamefont {{International Atomic
  Energy Agency, International Nuclear Structure and Decay Data Network.
  https://www-nds.iaea.org/}}},\ }\href@noop {} {\ }\BibitemShut {NoStop}%
\bibitem [{\citenamefont {Ingalls}(1976)}]{Ingalls:1976zz}%
  \BibitemOpen
  \bibfield  {author} {\bibinfo {author} {\bibfnamefont {P.~D.}\ \bibnamefont
  {Ingalls}},\ }\href {\doibase 10.1103/PhysRevC.14.254} {\bibfield  {journal}
  {\bibinfo  {journal} {Phys. Rev. C}\ }\textbf {\bibinfo {volume} {14}},\
  \bibinfo {pages} {254} (\bibinfo {year} {1976})}\BibitemShut {NoStop}%
%%CITATION = PHRVA,C14,254;%%
\bibitem [{\citenamefont {Singh}\ and\ \citenamefont
  {Evans}(1971)}]{SINGH1971475}%
  \BibitemOpen
  \bibfield  {author} {\bibinfo {author} {\bibfnamefont {B.}~\bibnamefont
  {Singh}}\ and\ \bibinfo {author} {\bibfnamefont {H.}~\bibnamefont {Evans}},\
  }\href {\doibase http://dx.doi.org/10.1016/0029-554X(71)90249-7} {\bibfield
  {journal} {\bibinfo  {journal} {Nuclear Instruments and Methods}\ }\textbf
  {\bibinfo {volume} {97}},\ \bibinfo {pages} {475 } (\bibinfo {year}
  {1971})}\BibitemShut {NoStop}%
\bibitem [{\citenamefont {Itzykson}\ and\ \citenamefont
  {Zuber}(2006)}]{itzykson2006quantum}%
  \BibitemOpen
  \bibfield  {author} {\bibinfo {author} {\bibfnamefont {C.}~\bibnamefont
  {Itzykson}}\ and\ \bibinfo {author} {\bibfnamefont {J.-B.}\ \bibnamefont
  {Zuber}},\ }\href@noop {} {\emph {\bibinfo {title} {Quantum field theory}}}\
  (\bibinfo  {publisher} {Courier Corporation},\ \bibinfo {year}
  {2006})\BibitemShut {NoStop}%
\end{thebibliography}%

\end{document}